# Giant nonlinear optical chirality in twisted heterobilayers

Xiang Zhang,[1]† Bo Li[2]†, Leyi Zhao,[1] Pengzhi Wang,[1] Luwei Zhou,[1] Jiangbo Peng,[1] Gan Wang,[2] Kian Ping Loh,[3] Tao-Yuan Du,[2 *] Mingjie Li,[1, 4, 5 *]

**Affiliations:**

[1]Department of Applied Physics, The Hong Kong Polytechnic University, Hung Hom, Kowloon, Hong Kong, China.

[2]School of Mathematics and Physics, China University of Geosciences, Wuhan 430074, China

[3]Department of Chemistry, National University of Singapore, Singapore, Singapore

[4]Photonics Research Institute, The Hong Kong Polytechnic University, Hung Hom, Kowloon, Hong Kong, China

[5]Shenzhen Research Institute, The HongKong Polytechnic University, Shenzhen, Guangdong 518057, China.

*Corresponding author. Email: ming-jie.li@polyu.edu.hk; duty@cug.edu.cn

†These authors contributed equally to this work

**Abstract:** Twisting two dissimilar monolayer semiconductors induces structural chirality that remains largely elusive in linear optics but becomes remarkably pronounced in the nonlinear regime. Here we demonstrate that $MoS_2/WSe_2$ heterobilayers exhibit giant, twist-tunable nonlinear chirality in second-harmonic generation (SHG). The sign of SHG circular dichroism is governed by structural handedness, and its magnitude reaches 1.96 near a 30° twist angle under 1260-nm excitation, approaching the theoretical limit of 2. Furthermore, reversed chirality is observed when light is incident from opposite directions. Using a layer-resolved model, we attribute this phenomenon to helicity-dependent interference between the two monolayer SHG fields, mediated by a nonlinear Pancharatnam-Berry phase. These findings establish that the relative orientation of atomically thin layers can deterministically control nonlinear chiral responses, identifying twisted 2D heterostructures as a versatile platform for nonlinear chiral photonics, frequency conversion, and ultracompact light-matter interfaces.

## Introduction

Chiral nonlinear optical responses provide a powerful route to control chiral light–matter interactions beyond the linear regime, driving growing interest in applications such as chiral nonlinear photonics [1-4], quantum nonlinear chiral devices [5-7], chiral nonlinear sensors [8, 9], By converting structural handedness into helicity-dependent optical asymmetry, non-centrosymmetric systems offer a promising foundation for chiral nonlinear optics [10-12]. Considerable effort has therefore been devoted to enhancing nonlinear chiral responses in a range of platforms, including engineered chiral metasurfaces [3, 13-16], chiral perovskites [17-19], coherently stacked boron nitride nanotubes [20], nanoscrolled molybdenum disulfide [21], and other low-dimensional chiral materials [22, 23]. Despite this progress, existing platforms remain limited by modest nonlinear chirality, environmental instability, or complex nanofabrication, underscoring the need for a simple structurally tunable system capable of producing giant nonlinear chiral responses.

Two-dimensional (2D) materials provide an attractive opportunity to overcome these limitations because their superior optical properties can be engineered through both composition and stacking geometry [24-26]. Transition-metal dichalcogenides (TMDs), in particular, exhibit strong second-order nonlinear susceptibilities and symmetry-governed optical selection rules, making them a versatile platform for nonlinear photonics [27-29]. When assembled into van der Waals heterostructures, these materials gain an additional structural degree of freedom—the interlayer twist angle—which can profoundly reshape their electronic and optical behavior [29, 30]. Early studies of twisted TMDs bilayers showed that second-harmonic generation (SHG) under linearly polarized excitation is governed by coherent superposition of the nonlinear fields emitted by the individual layers [31, 32]. In parallel, structurally chiral atomically thin systems were realized in twisted bilayer graphene and in multilayer twisted TMDs assemblies with controllable chirality [33, 34]. However, these advances primarily focused on SHG interference, interlayer coupling, or structural chirality itself. Nonlinear chiral responses in twisted 2D heterobilayers have not yet been investigated, and it remains unclear whether and how interlayer twist can be used as a direct and general control knob for producing giant nonlinear chirality in such systems.

Here we demonstrate that the interlayer twist angle in $MoS_2$/$WSe_2$ heterobilayers provides precisely such a knob. We observe that interlayer twist induces giant nonlinear chirality that is nearly silent in linear optics yet becomes dominant in the nonlinear regime, with SHG-circular-dichroism (SHG-CD) reaching 1.96 near a 30° twist angle under 1260-nm excitation. We further show that this near-limit response originates from a helicity-dependent nonlinear Pancharatnam–Berry (PB) phase that governs the coherent interference between the second-harmonic fields emitted by the two monolayers. The same framework explains the measured wavelength dependence of SHG-CD and provides a simple design principle for tailoring nonlinear chiroptical responses in atomically thin van der Waals heterostructures and highlighting the heterobilayers as a compelling platform for ultracompact chiral photonic functionality.

## Results

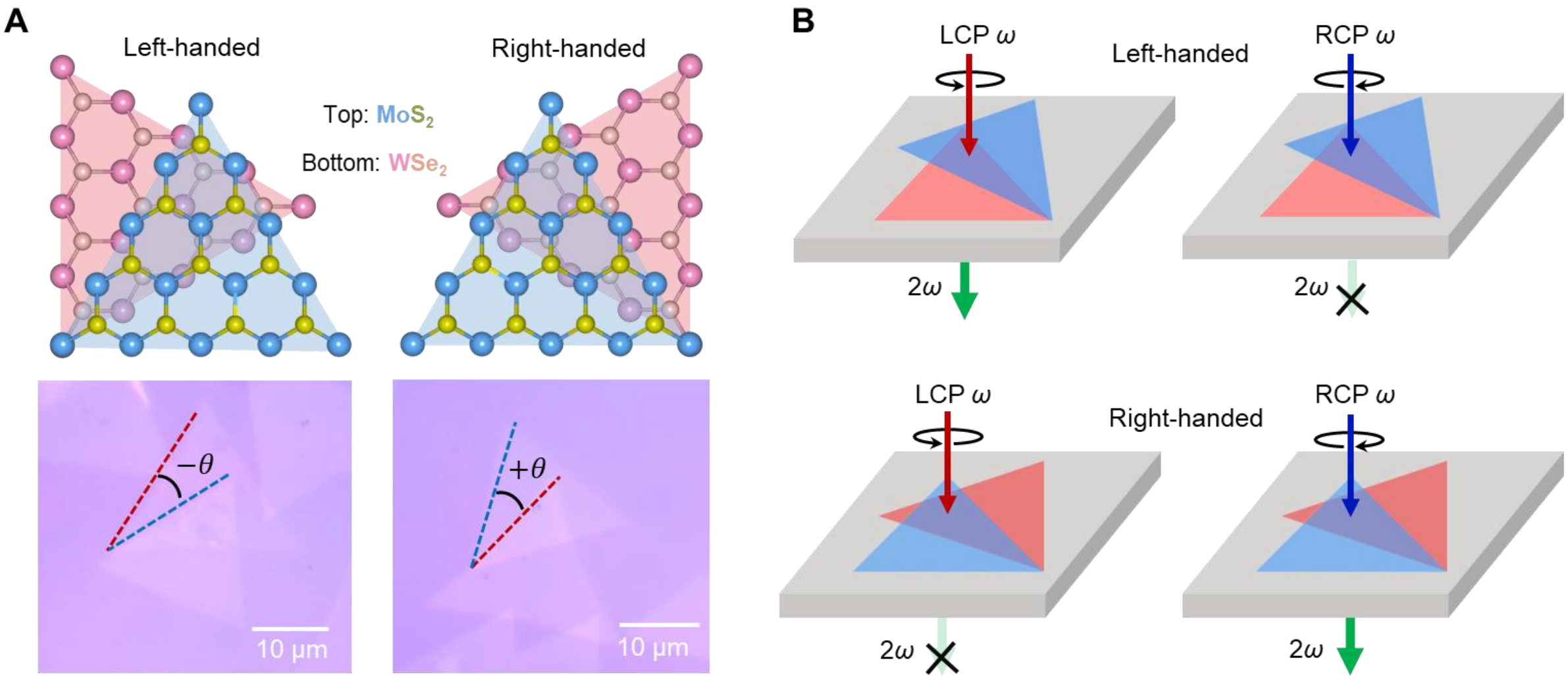

**Fig. 1. Nonlinear chirality in a twisted heterobilayer.** (**A**) Schematic illustrations of left-handed and right-handed $MoS_2/WSe_2$ heterobilayers. Blue and pink triangles represent $MoS_2$ and $WSe_2$ monolayers, respectively. Optical micrographs of representative left-handed (-) and right-handed (+) heterostructures with twist angles near 30° are shown below. (**B**) Schematic illustration of the chiral SHG response of left-handed and right-handed heterobilayers under left-circularly polarized (LCP) and right-circularly polarized (RCP) excitation. Depending on the handedness of the heterobilayer, the pump helicity, and the twist angle, the SHG signal can be either enhanced or strongly suppressed, giving rise to helicity-selective nonlinear optical response.

### Twisted nonlinear chiral heterobilayers

By controlling the relative twist angle between monolayer $MoS_2$ and $WSe_2$, the mirrored left- and right-handed chiral heterobilayers can be generated, as illustrated in Fig. 1A. In our structures, $MoS_2$ forms the top layer and $WSe_2$ the bottom layer. Optical micrographs of representative left- and right-handed $MoS_2/WSe_2$ heterobilayers synthesized by chemical vapor deposition confirm the formation of the stacked structures. We define the bottom $WSe_2$ layer as the reference orientation and the twist angle $\theta$ as the in-plane rotation of the top $MoS_2$ layer, where $\theta$ = 0° corresponds to AA stacking. Under this convention, a negative $\theta$ (clockwise rotation) corresponds to the experimentally defined left-handed heterobilayer, while positive $\theta$ (anti-clockwise rotation) corresponds to the right-handed one. The handedness of the excitation beam follows the standard convention defined by the rotation of the electric field for light propagating toward the viewer.

We first examined the linear optical response of heterobilayers with twist angles near 30°. Linear absorption and circular dichroism measurements show no measurable difference between left- and right-circularly polarized excitation (Note S1 and fig. S1A,B), indicating that the linear chiral response of this two-layer system is negligibly small. That result contrasts sharply with the nonlinear regime and underscores the usefulness of SHG-CD as a highly sensitive probe of structural chirality in atomically thin materials [22]. Fig. 1B summarizes this central idea: although the heterobilayer is nearly silent in linear circular dichroism, it can respond strongly and selectively in second-harmonic generation (SHG).

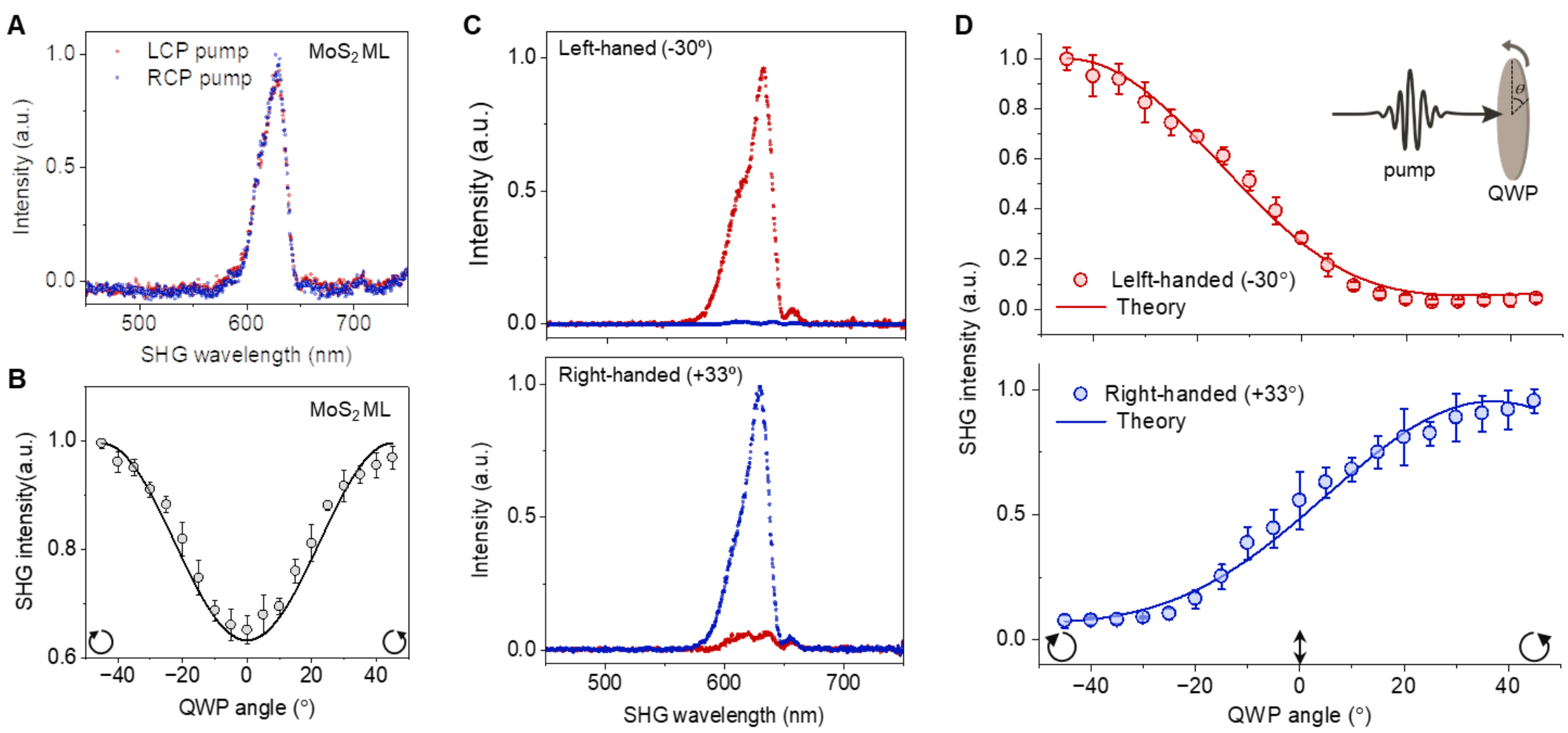

**Fig. 2. Helicity-controlled SHG in twisted heterobilayer.** (**A**) SHG spectra of a monolayer $MoS_2$ under 1260-nm LCP and RCP excitation. (**B**) Ellipticity dependence of the SHG peak intensity for monolayer $MoS_2$. Solid lines denote the corresponding theoretical calculations (Notes S2–S3). (**C**) SHG spectra of left-handed and right-handed $MoS_2/WSe_2$ heterobilayers under LCP and RCP excitation. (**D**) Ellipticity dependence of the SHG peak intensities of left-handed and right-handed $MoS_2/WSe_2$ heterobilayers under LCP and RCP excitation. Solid

lines denote the corresponding theoretical calculations (Notes S2–S3). Error bars represent the standard deviation of SHG intensities from three independent measurement sets, and the plotted data points indicate the corresponding mean values.

**Pump ellipticity-controlled SHG**

SHG in monolayer TMDs arises from the absence of inversion symmetry in their hexagonal lattices [35, 36]. Under left- or right-circularly polarized pumping, the SHG intensities from an isolated monolayer are nearly identical. A single monolayer belongs to a non-centrosymmetric point group and therefore supports a strong in-plane second-order nonlinear susceptibility [37, 38]. At the same time, the presence of a horizontal mirror plane renders the monolayer structurally achiral [39]. Accordingly, under normal-incidence circularly polarized excitation, the SHG intensity of monolayer $MoS_2$ and $WSe_2$ is essentially independent of pump handedness (Fig. 2A and fig. S2). As the pump polarization evolves from linear to circular, the SHG signal increases smoothly with ellipticity (Fig. 2B), consistent with the expected polarization dependence of monolayer SHG (Note S3).

The heterobilayer behaves qualitatively differently. In the left-handed sample, under 1260-nm excitation, the SHG intensity is maximal for left-circularly polarized excitation and minimal for right-circularly polarized excitation (Fig. 2C, top). In the right-handed sample, the trend is reversed (Fig. 2C, bottom). We quantify the nonlinear circular dichroism by the SHG dissymmetry factor

$$g_{\mathrm{SHG-CD}} = 2(I_{2\omega}^{L} - I_{2\omega}^{R})/(I_{2\omega}^{L} + I_{2\omega}^{R}), \quad (1)$$

where $I_{2\omega}^{L}$ and $I_{2\omega}^{R}$ are the SHG intensities under LCP and RCP light pumping, respectively. From the spectral peaks in Fig. 2C, $g_{\mathrm{SHG-CD}}$ reaches 1.96 for the left-handed sample near -30° and 1.85 for the right-handed sample near +33°, both close to the theoretical maximum of 2.

The dependence on pump ellipticity further emphasizes the helicity selectivity of the twisted heterobilayer. For the left-handed structure, the SHG intensity decreases monotonically as the pump is varied from left- to right-circular polarization. For the right-handed structure, the signal evolves in the opposite direction (Fig. 2D). The mirror-opposite polarization dependence of the two enantiomers shows that substantial nonlinear circular dichroism emerges only after two individually achiral monolayers are combined into a twisted, structurally chiral bilayer.

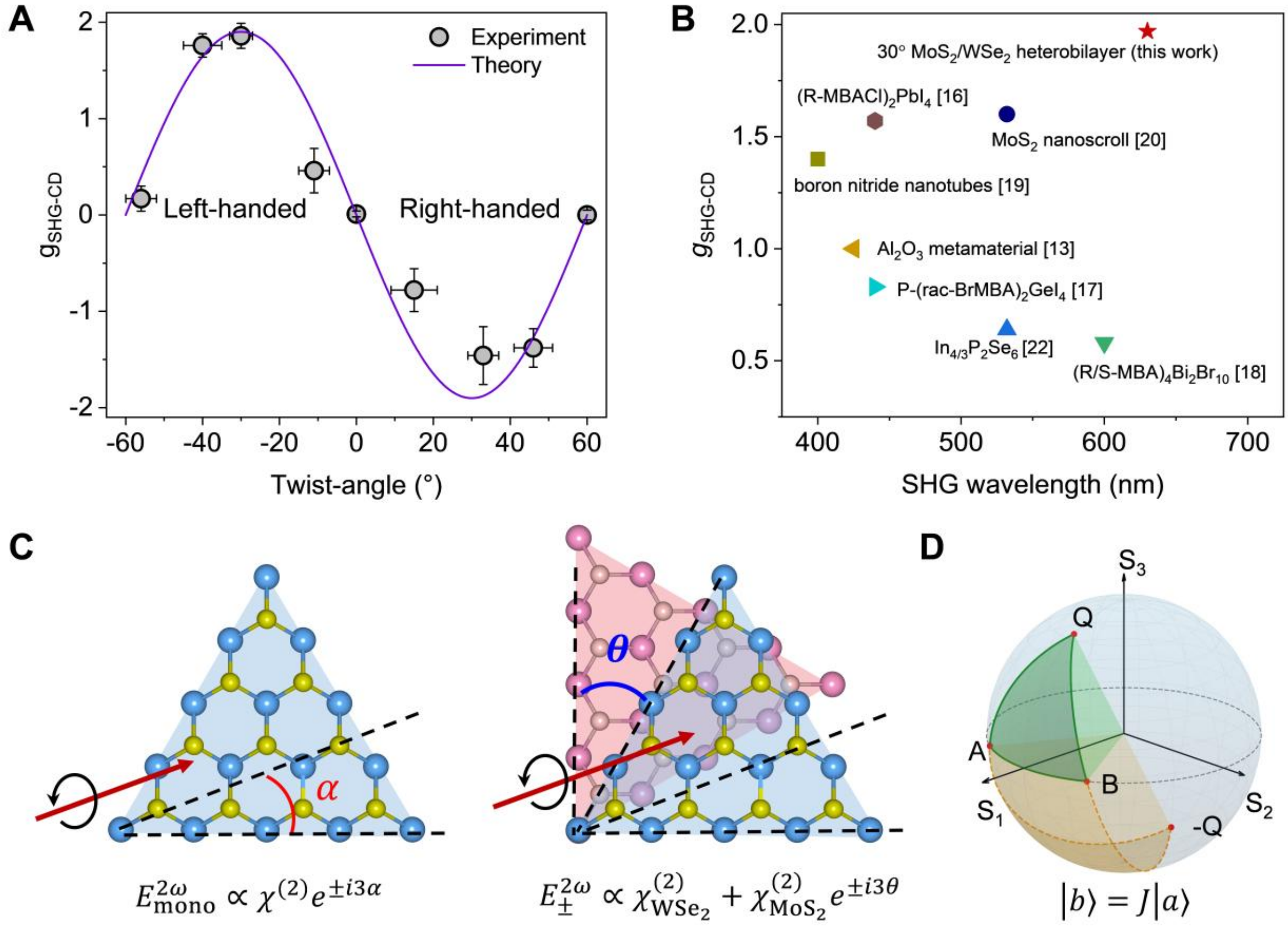

**Fig. 3. SHG-CD and nonlinear Pancharatnam-Berry phase interference. (A)** Experimental (dots) and calculated (solid lines) SHG-CD under 1260-nm pump as a function of twist angles. Error bars represent the standard deviation of SHG intensities and twist angles from three independent measurement sets; data points are the corresponding mean values. **(B)** Comparison of the maximum $g_{\text{SHG-CD}}$ with values reported for other materials. (**C**) Schematic illustration of coherent interference between layer-resolved SHG fields in a twisted $MoS_2/WSe_2$ heterobilayer. Reversing the pump helicity changes the relative phase and switches the interference condition, producing SHG-CD. (**D**) Geometric interpretation of twist-induced nonlinear Pancharatnam–Berry phase on the Poincaré sphere (Note S4). Points A and B represent the second-harmonic generation (SHG) polarization states of the fixed bottom $WSe_2$ monolayer and the twisted top $MoS_2$ monolayer, respectively. The two states are related by the twist-induced nonlinear optical rotation operator J, which is uniquely determined by the interlayer twist angle θ, satisfying $|b\rangle = J|a\rangle$.

**Twist-angle dependence of SHG-CD**

The twist-angle dependence of $g_{\text{SHG-CD}}$ is summarized in Fig. 3A. The response is strongly suppressed at the high-symmetry angles of -60°, 0°, and 60°, but rises sharply to a pronounced maximum near 30°. This behavior follows the general symmetry principle that chirality in stacked 2D van der Waals layers vanishes at interlayer angles of 360°/2$n$ for crystals with $n$-fold rotational symmetry [40]. Because both $MoS_2$ and $WSe_2$ possess threefold rotational symmetry, the heterobilayer becomes effectively achiral when the twist angle is a multiple of 60°. Near 30°, by contrast, the structure is maximally nonlinear chiral and the SHG-CD is correspondingly strongest.

The magnitude of the response is notable. The maximum $g_{\mathrm{SHG-CD}}$ observed here exceeds values reported for a range of other nonlinear chiral systems (Fig. 3B) [13, 17-21, 41], identifying twisted $MoS_2/WSe_2$ heterobilayers as a compelling platform for compact chiral photonic functionality.

To assess the generality of the phenomenon, we further examined an artificially stacked 21° $MoS_2/WSe_2$ heterobilayer assembled from exfoliated crystals, as well as a twisted $MoS_2/WS_2$ heterobilayer. In both cases, the supplementary data show the same qualitative behavior (fig. S3 and S5): a pronounced helicity-dependent SHG response that reverses with handedness and remains substantial away from the specific $MoS_2/WSe_2$ samples highlighted. These controls indicate that giant SHG-CD is not an isolated feature of a single specimen, but a broader consequence of combining non-centrosymmetric monolayers into a twisted chiral heterobilayer.

**Pancharatnam-Berry phase interference**

The physical origin of the helicity-dependent SHG is coherent interference between the nonlinear fields emitted by the two monolayers (Fig. 3C). For an isolated monolayer, LCP ($\sigma = +1$) or RCP ($\sigma = -1$) excitation contributes only an overall phase factor $e^{i\sigma\alpha}$ to the SHG field and therefore does not alter the intensity [42, 43]. In a twisted heterobilayer, however, rotating one $D_{3h}$ monolayer by an angle $\theta$ imprints a helicity-dependent nonlinear Pancharatnam–Berry (PB) phase of $e^{\pm i3\theta}$ on its coherent second-harmonic field (Note S3) [44-47]. Fig. 3D provides a geometric interpretation on the Poincaré sphere. This phase enters the relative phase between the two layer-resolved SHG amplitudes. Reversing the pump helicity flips the sign of the PB phase ($+3\theta \leftrightarrow -3\theta$), thereby switching the interference condition between more constructive and more destructive superposition. The resulting interference contrast gives rise directly to SHG-CD. The $3\theta$ dependence naturally produces a 60° periodicity, and reversing the sign of the twist angle reverses the sign of $g_{\mathrm{SHG-CD}}$.

The total SHG field of the heterobilayer can be written as the coherent sum of the two layer-resolved nonlinear responses. In the ultrathin limit relevant to twisted TMD heterobilayers, each monolayer is far thinner than both the optical wavelength and the SHG coherence length, so propagation phases accumulated within an individual layer are negligible to leading order. The dominant physics is therefore not conventional phase matching inside a bulk medium, but coherent interference between two deeply subwavelength nonlinear dipole sources [48, 49]. Taking the bottom layer as the reference and defining $\theta$ as the relative rotation of the top layer, we can write the helicity-dependent SH field as (see Note S5 for the derivation):

$$E_{\mp,2\omega}=\frac{i\sqrt{2}\omega}{c\bar{n}(2\omega)}E_{\pm,\omega}^{2}\left(\tilde{\chi}_1+\tilde{\chi}_2e^{\pm i3\theta}\right) \tag{2}$$

Here, $E_{\pm,\omega}$ denotes the electric-field amplitude of the left- and right-circularly polarized fundamental wave, and $E_{\mp,2\omega}$ denotes the electric-field amplitude of the oppositely handed SH wave. $\bar{n}(2\omega)$ is the effective refractive index at the SH frequency. $\tilde{\chi}_1$ and $\tilde{\chi}_2$ represent the effective 2D second-order nonlinear susceptibilities of the two layers; (the superscript "2D" is omitted here).

This compact expression makes clear that the twist enters solely through the helicity-dependent nonlinear PB phase.

This ultrathin-limit treatment does not imply that interlayer coupling is unimportant. On the contrary, charge transfer, excitonic hybridization, photoluminescence quenching, and resonance shifts are all folded into the effective complex nonlinear coefficients of the two monolayers, including their amplitudes, intrinsic phases, and spectral dependence. Within this framework, interlayer coupling renormalizes the two layer responses, whereas the dominant mechanism responsible for the giant helicity contrast is the PB-phase-controlled interference between them.

From the SHG intensity $I_{2\omega}(\sigma) = |E|^2 \propto \left|\tilde{\chi}_1 + \tilde{\chi}_2e^{\pm i3\theta}\right|^2$, the dissymmetry factor can accordingly be expressed as. (see Note S6 for the derivation):

$$g_{\mathrm{SHG-CD}}(\omega,\theta) = \frac{-4\rho_1(\omega)\rho_2(\omega)\sin\Delta\phi_{\mathrm{eff}}(\omega)\sin 3\theta}{\rho_1^2(\omega)+\rho_2^2(\omega)+2\rho_1(\omega)\rho_2(\omega)\cos\Delta\phi_{\mathrm{eff}}(\omega)\cos 3\theta}. \tag{3}$$

where the $\Delta\varphi_{\mathrm{eff}}(\omega) = \varphi_2(\omega) - \varphi_1(\omega)$ is the intrinsic phase difference between the two monolayer nonlinear responses. The $\rho_1(\omega)$ and $\rho_2(\omega)$ are amplitude of frequency-dependent complex susceptibilities $\chi_1(\omega)$ and $\chi_2(\omega)$. The calculated $g_{\mathrm{SHG-CD}}(\theta)$ from Eq. (3) reproduces the measured angular dependence well (Fig. 3A).

Because the PB phase changes sign with both pump helicity and sample handedness, the model naturally captures the enantiomeric symmetry observed experimentally. At $\theta = 0°$ and symmetry-equivalent angles, the heterobilayer is effectively achiral and $g_{\mathrm{SHG-CD}}$ vanishes.

The particularly large response near $\theta \approx 30°$ can be understood directly from the phase relation. At this twist angle, the helicity-dependent phase difference between the two-layer contributions approaches a quarter-cycle offset, maximizing the interference contrast and driving $g_{\mathrm{SHG-CD}}$ toward its theoretical limit. In that regime, the two complex nonlinear amplitudes become nearly orthogonal in the complex plane, producing the extreme helicity selectivity seen in experiment. Small deviations between experiment and theory are plausibly attributable to twist-angle disorder, local strain, or other sample imperfections.

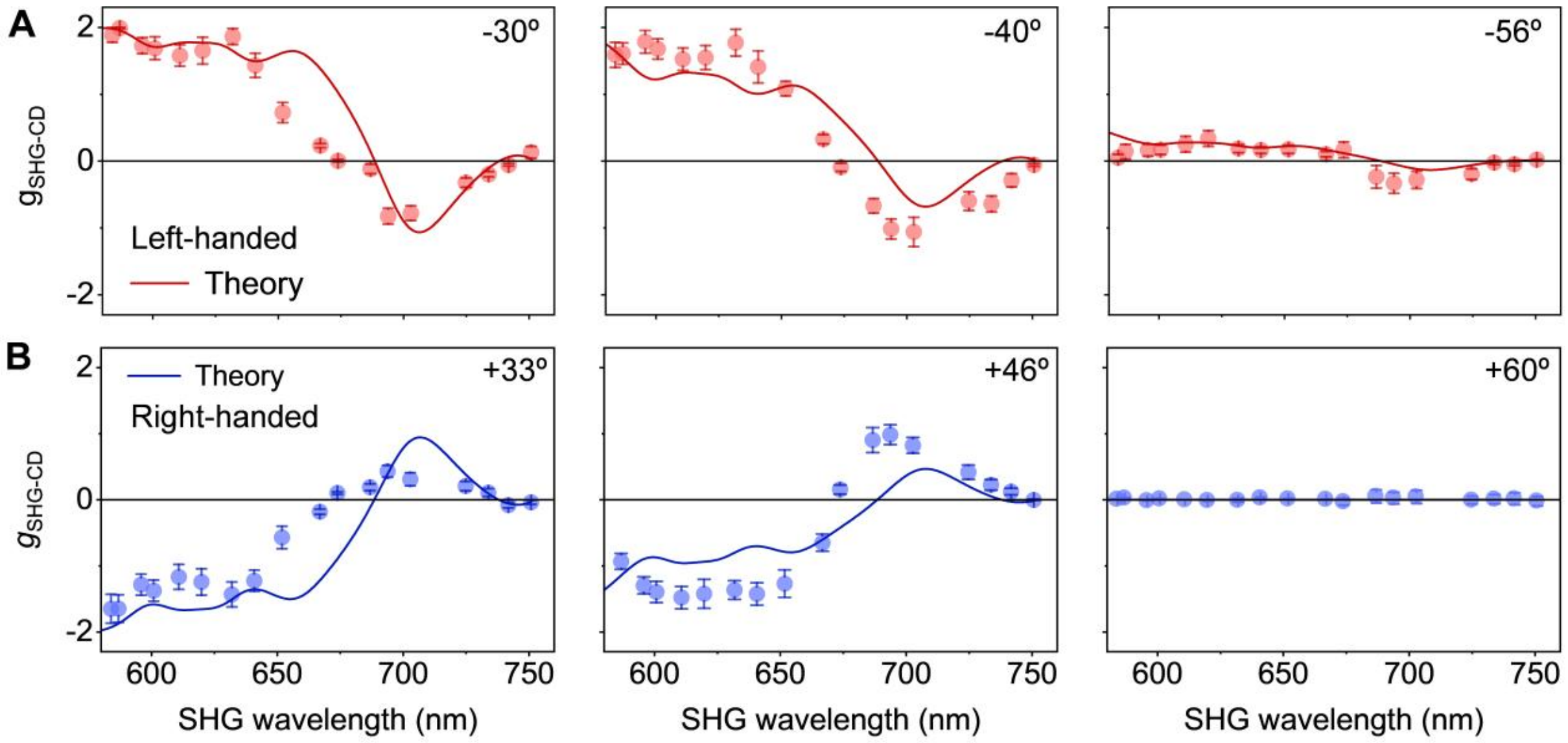

**Fig. 4. Wavelength-dependent SHG-CD.** Experimental (dots) and calculated (solid lines) SHG-CD spectra for a series of twisted $MoS_2/WSe_2$ heterobilayers measured over the investigated pump-wavelength range. Error bars represent the standard deviation of SHG intensities from three independent measurement sets, and the plotted data points denote the corresponding mean values. The wavelength dependence reflects the combined effects of the twist-induced geometric phase and the frequency-dependent complex nonlinear susceptibilities of the constituent monolayers.

## Wavelength dependent SHG-CD

Fig. 4 shows the measured wavelength dependence of SHG-CD for a series of $MoS_2/WSe_2$ heterobilayers. In our model, the geometric PB phase $3\theta$ is fixed by the structure and therefore does not depend on wavelength. The spectral evolution of $g_{\mathrm{SHG-CD}}$ instead arises from the dispersive complex nonlinear responses of the two constituent monolayers $A_1(\omega)$, $A_2(\omega)$, and $\Delta\phi_{\mathrm{eff}}(\omega)$: as the pump wavelength changes, so do the amplitudes and relative intrinsic phase of their effective nonlinear susceptibilities, whose frequency-dependent complex $\chi^{(2)}(\omega)$ are obtained from the so-called length-gauge formalism developed by Aversa and Sipe et al [50, 51] (see Note S6 for the derivation) .The relative phase between the two layer-resolved SHG sources therefore evolves continuously with wavelength, changing the helicity-dependent interference condition.

This picture explains the observed zero crossings and sign reversals of SHG-CD. For a fixed twist angle $\theta \neq 0, \pm 60°$, $g_{\text{SHG-CD}}(\omega_c, \theta) \approx 0$ when: $\sin\left(\Delta\varphi_{\text{eff}}(\omega_c)\right) \approx 0$, i.e., when $\Delta\phi_{\text{eff}}(\omega_c) = m\pi$ for integer $m$. When the wavelength is tuned across $\omega_c$, $\sin(\Delta\phi_{\text{eff}}(\omega))$ changes sign, and the SHG-CD correspondingly flips sign. This explains the experimentally observed sequence with increasing wavelength: for a fixed twist angle, the dissymmetry factor vanishes when the wavelength-dependent relative phase compensates the twist-induced geometric phase, collapsing the contrast between the two helicity channels. As the wavelength is tuned further, the relative phase passes through this compensation condition and the sign of SHG-CD reverses. The experimentally observed sequence—a finite response that decreases to zero, changes sign, and then trends back toward zero at longer wavelength—thus follows directly from interference between a fixed, twist-controlled geometric phase and a dispersive, material-dependent phase difference.

The wavelength dependence therefore does not reflect any spectral variation of the PB phase itself. Rather, it emerges from the interplay between a geometrically imposed helicity asymmetry and the frequency-dependent complex nonlinear response of each layer. Because our treatment is both layer resolved and frequency dependent, and uses the *ab initio* complex $\chi^{(2)}(\omega)$ of $MoS_2$ and $WSe_2$ as material input, it quantitatively captures the measured spectral trends. More broadly, it provides a practical design rule: the twist angle determines the handedness and geometric phase, while the dispersive monolayer susceptibilities determine where SHG-CD is enhanced, suppressed, or changes sign. Residual differences in spectral shape and magnitude likely arise from the extreme phase sensitivity of interlayer SHG interference in real heterobilayers.

**Conclusion**

We have shown that interlayer twist engineering in van der Waals heterostructures provides a simple and effective route to giant nonlinear chiroptical responses. Twisted 2D TMDs heterobilayers exhibit pronounced second-harmonic-generation circular dichroism whose sign is set by structural handedness and whose magnitude depends strongly on twist angle, reaching 1.96 near 30° under 1260-nm excitation. Similar behavior in additional heterobilayers indicates that the effect is not unique to a single material pair. These results establish nonlinear twistronics as a promising platform for compact, tunable chiral photonic functionalities.

We further developed a helicity-dependent nonlinear Pancharatnam-Berry phase interference model that explicitly links interlayer twist to SHG-CD. In the ultrathin limit appropriate for atomically thin van der Waals heterostructures, the problem reduces to coherent superposition of two effective complex nonlinear amplitudes, with the twist-induced $\pm 3\theta$ geometric phase controlling the helicity dependence of their interference. Beyond reproducing the experimental angular and spectral trends, this framework identifies the condition for approaching the theoretical limit $g_{\text{SHG-CD}} = 2$ and provides a general basis for understanding and engineering nonlinear chiroptical responses in twisted van der Waals heterostructures.

**Acknowledgments**

**Funding:** M.L. acknowledges the fund support from Research Grant Council of Hong Kong (Project No. PolyU 25301522, PolyU 15301323, PolyU 15300824, C5003-24E), National Natural Science Foundation of China (22373081), the Shenzhen Science, Technology and Innovation Commission (JCYJ20210324131806018), GuangDong Basic and Applied Basic Research Foundation (2024A1515011261). KP. Loh acknowledges the fund support by Hong Kong Jockey Club STEM lab of 2D Quantum Material (2023-2033).

**Author contributions:**

Experiments: X. Z., L. Y. Z., P. Z. W., L. W. Z and J. B. P

Experimental data analysis: M. J. L., T. Y. D, X. Z. and B. L.
Numerical analysis: T. Y. D. led the theoretical analyses of the PB phase interference; B. L., G. W. and T. Y. D. together performed the numerical simulations.

Idea and design of experiment: X. Z. and M. J. L.

Writing – original draft: X. Z., B. L., T. Y. D. and M. J. L

Writing – review & editing: M. J. L., T. Y. D. and K. P. L

**Competing interests:** The authors declare no competing interests.